\documentclass{emulateapj}
\usepackage{color}

\slugcomment{Received 2013.12.31; Accepted}

\shorttitle{SGRB Rate as Gravitational Wave Source}
\shortauthors{}

\begin{document}


\title{Short Gamma Ray Burst  Formation Rate from 
BATSE data using $E_p-L_p$ correlation and the minimum gravitational wave event rate of coalescing compact binary }


\author{
Daisuke Yonetoku\altaffilmark{1},
Takashi Nakamura\altaffilmark{2},
Tatsuya Sawano\altaffilmark{1},
Keitaro Takahashi\altaffilmark{3}, 
and 
Asuka Toyanago\altaffilmark{1},}
\email{yonetoku@astro.s.kanazawa-u.ac.jp}
\email{takashi@tap.scphys.kyoto-u.ac.jp}
\altaffiltext{1}{College of Science and Engineering, 
School of Mathematics and Physics,
Kanazawa University, Kakuma, Kanazawa, Ishikawa 920-1192, Japan}
\altaffiltext{2}{Department of Physics, Kyoto University, Kyoto
606-8502, Japan}
\altaffiltext{3}{Faculty of Science, Kumamoto University, Kurokami,
Kumamoto, 860-8555, Japan}

\begin{abstract}
Using 72 Short Gamma Ray Bursts (SGRBs) with well determined spectral 
data observed by BATSE, we determine their redshift and the luminosity 
by applying $E_p$--$L_p$ correlation for SGRBs found by \cite{tsutsui13}.
For 53 SGRBs with the observed flux brighter than
$4 \times 10^{-6}~{\rm erg~cm^{-2}s^{-1}}$, the cumulative redshift
distribution up to $z=1$ agrees well with that of 22  {\it Swift}~SGRBs.
This suggests that the redshift determination 
by the $E_p$--$L_p$ correlation for SGRBs works  well. 
The minimum event rate at $z=0$ is estimated as
$\rho_{SGRB}(0) = 6.3_{-3.9}^{+3.1} \times 10^{-10}~{\rm events~Mpc^{-3}yr^{-1}}$
so that the minimum beaming angle is $0.6^\circ-7.8^\circ$ assuming
the merging rate of 
$10^{-7}-4\times 10^{-6}~{\rm events~Mpc^{-3}yr^{-1}}$ suggested
from the binary pulsar data. Interestingly, this angle is consistent
with that for SGRB130603B of $\sim 4^\circ-8^\circ$\citep{fong13b}.
On the other hand, if we assume the beaming angle of $\sim 6^\circ$
suggested from four SGRBs with the observed value  of beaming angle, the minimum event rate including off-axis 
SGRBs is estimated as
$\rho_{SGRB,all}^{min}(0)=1.15_{-0.71}^{+0.57}\times 10^{-7}~{\rm events~Mpc^{-3}yr^{-1}}$.
If SGRBs are induced by coalescence of binary neutron stars (NSs) and/or
black holes (BHs), this event rate leads to the minimum gravitational-wave
detection rate of $\rm 3.9_{-2.4}^{+1.9} (152_{-94}^{+75})~events~y^{-1}$ for NS-NS (NS-BH) binary,
respectively, by a worldwide network with KAGRA, advanced-LIGO, 
advanced-Virgo, and GEO.
\end{abstract}



\section{Introduction}
Although the number of observed Short Gamma-Ray Bursts (SGRBs) is 
increasing, their central engine is still a big mystery. A major 
candidate is coalescence of compact objects such as neutron stars 
and stellar-mass black holes. One of the keys to confirm this idea 
is the formation rate of SGRBs as a function of redshift. In fact, 
if SGRBs are truly induced by coalescence of compact objects, 
the SGRB formation rate will track the star formation rate with 
some delay time. Further, if this is the case, SGRBs are expected 
to be accompanied by substantial gravitational-wave emission. 
Thus, the local SGRB formation rate is directly related to the 
expected number of gravitational-wave events for the next-generation 
gravitational-wave detectors such as 
KAGRA \footnote{http://gwcenter.icrr.u-tokyo.ac.jp/en/}, 
advanced-LIGO \footnote{http://www.ligo.caltech.edu/}, 
advanced-VIRGO \footnote{http://www.ego-gw.it/index.aspx/} and 
GEO \footnote{http://www.geo600.org/}.

However, the number of SGRBs with known redshift is very small 
($\sim 20$) so that the formation rate is not easy to estimate. 
Previous studies have estimated the formation rate assuming 
the functional form of the formation rate and the luminosity 
function and fitting the data to derive model parameters 
\citep{guetta05,guetta06,nakar06,dietz11,metzger12,petrillo13}. 
In this approach, the result depends on the functional form of 
the model and has large statistical errors due to the small number 
of SGRBs used to fit the model.

On the other hand, as to Long Gamma-Ray Bursts (LGRBs), 
the formation rate has been estimated much more precisely 
and robustly. This is because the correlation between 
the spectral peak energy and luminosity was found and used to 
estimate the redshift of LGRBs without known redshift. 
First, \cite{yonetoku04} analyzed data of 16 LGRBs observed 
by {\it CGRO} BATSE and {\it BeppoSAX} with known redshifts, 
and found the $E_p-L_p$ correlation between the $E_p$ and 
the peak luminosity $L_p$ integrated for 1~s time interval 
at the peak as,
\begin{equation}\label{eq:lgrbeplp}
L_p=\rm 2.34\times10^{51}erg~s^{-1}(\frac{E_p}{100keV})^2.
\end{equation}
The linear correlation coefficient of the $\log E_p-\log L_p$ 
correlation is 0.958 and the chance probability is 
$5.31 \times 10^{-9}$. 
Then \cite{ghirlanda05a, ghirlanda05b, krimm09, yonetoku10} checked 
the properties of the correlation and confirmed its reliability. 
Using this correlation, \cite{yonetoku04} estimated the redshift 
for 689 bright BATSE LGRBs without known redshift and derived 
the luminosity function and the formation rate.

As to SGRBs, however, due to the small number of the events with 
known redshifts and the good spectra to determine $E_p$, it has been 
difficult to do a similar analysis. Recently, \cite{tsutsui13} 
succeeded in determining $E_p-L_p$ correlation for SGRBs. 
They used 8 secure SGRBs out of 13 candidates and obtained 
\begin{equation}\label{eq:sgrbeplp}
L_p = \rm 7.5 \times 10^{50}erg~s^{-1}(\frac{E_p}{100keV})^{1.59},
\end{equation}
where $E_p$ is from the time integrated spectrum again while $L_p$ 
was taken as the luminosity integrated for 64~ms time intervals 
at the peak considering the shorter duration of SGRB. The linear 
correlation coefficient of the $E_p-L_p$ correlation is 0.98 and 
the chance probability is $1.5 \times 10^{-5}$. Although this is 
not so tight as that for LGRBs due to the fact that the number of 
SGRBs is half of that of LGRBs, it is accurate enough to use as 
a redshift indicator for many SGRB events without known redshifts.

In this article, we determine the redshifts of SGRBs observed by 
BATSE using the $E_p$--$L_p$ correlation mentioned above. 
Then, we obtain a non-parametric estimate of the luminosity 
function and SGRB formation rate versus redshift based on much more 
samples compared with previous studies. This article is organized 
as follows. In \S 2 , we describe the observations and data analyses. 
After that, we show the redshifts estimated by $E_p$--$L_p$ correlation 
for SGRBs, and obtain the cumulative redshift distribution and 
compare it with the observed one. We also show the cumulative 
luminosity function and the SGRB formation rate as a function 
of redshift with non-parametric method, i.e. without any assumptions
on both distributions. \S 3 is devoted to discussions and 
implications of the results.

\section{Observations \& Data Analyses}
\label{sec:analyses}
\subsection{Data Selection}
\label{sec:data}
We used the BATSE current burst catalog which contains 2704 GRBs 
during its life time ($\sim 9.2$ years) in orbit. The average fraction 
of sky coverage of BATSE instruments is 0.483, so the effective life
time is $\sim 4.4$~years for the entire sky. 
We selected events with the short time duration of $T_{90} < 2$~sec 
in the observer frame as SGRB candidates. Here $T_{90}$ is measured 
as the duration of the time interval during which 90~\% of the total 
observed counts have been detected. After that, we selected
the brightest 100~SGRBs in 1,024~ms peak photon flux. 
The peak flux of all events we selected is brighter than 
$1~{\rm photons~cm^{-2}s^{-1}}$. The trigger efficiency of BATSE 
instrument is almost 100~\% (larger than 99.988~\%), so we 
can estimate SGRB rate without any correction in this point.

\subsection{Spectral Analyses}
We used spectral data detected by the BATSE LAD detectors and 
performed the standard data reduction method. Then we succeeded in 
analyzing spectral data for 72~events. The other 28 events are 
statistically poor or the variable background condition, so we 
failed to obtain spectral parameters for the standard analyses.

We used the spectral model of the smoothly broken power-law model,
so called Band function \citep{band93} as following;
\begin{eqnarray}\label{eq:band}
N(E) = \left\{ 
\begin{array}{ll}
        A \Bigl( \frac{E}{100~{\rm keV}} \Bigr)^{\alpha}
        \exp(- \frac{E}{E_{0}})\\ 
        {\rm for} \ E \le (\alpha - \beta) E_{0},\\
        A \Bigl( \frac{E}{100~{\rm keV}} \Bigr)^{\beta} \Bigl( 
        \frac{(\alpha - \beta) E_{0}}{100~{\rm keV}}\Bigr)^{\alpha - \beta} 
        \exp(\beta - \alpha) \\
        {\rm for} \ E \ge (\alpha - \beta) E_{0},
\end{array}
\right.
\end{eqnarray}
where, $N(E)$ is in unit of $\rm{photons~cm^{-2}s^{-1}keV^{-1}}$.
The spectral parameters of $\alpha$, $\beta$ and $E_{0}$ is 
the low- and high-energy index, and the energy at the spectral break,
respectively. For the case of 
$\beta < -2$ and $\alpha > -2$, the peak energy can be derived as 
$E_{p} = (2+\alpha) E_{0}$. In previous work, although 
\citet{ghirlanda09} performed spectral analyses for 79 SGRBs with 
cutoff power-law model, we used the Band function for all events 
in this work because the $E_p$--$L_p$ correlation by \cite{tsutsui13} 
is based on the Band function. If we can not determine the high energy 
spectral index $\beta$, we fixed the parameter as $\beta = -2.25$ 
which is average value of bright events.

\subsection{Redshift Estimation for SGRBs}
For long GRBs, there are well known correlations between $E_{p}$ and
brightness like Amati -- Yonetoku -- Ghirlanda correlations 
\citep[e.g.][]{amati02, yonetoku04, amati06, ggl04, yonetoku10}.
Recently \citet{tsutsui13} reported the $E_{p}$--luminosity correlation 
in SGRBs as equation~\ref{eq:sgrbeplp}.
This is $\sim 5$ times dimmer than the $E_{p}$--luminosity correlation 
of long GRBs (see equation~\ref{eq:lgrbeplp}).
This equation can be rewritten as 
\begin{eqnarray}
\frac{d_{L}^{2}}{(1+z)^{1.59}} = \frac{10^{50.88}}{4 \pi F_{p}}
\Bigl(\frac{E_{p}}{100~{\rm keV}} \Bigr)^{1.59}~~~{\rm erg~s^{-1}}.
\end{eqnarray}
Here, the right side of the equation is composed by observed values.
As \citet{yonetoku04} performed, using the observed $E_{p}$ and 
64~msec peak flux, we can estimate the pseudo redshift and
luminosity distance for each event. Then we used the cosmological 
parameters of $\Omega_{m}=0.3$, $\Omega_{\Lambda}=0.7$ and 
Hubble parameters of $H_{0}=71~{\rm km~s^{-1}Mpc^{-1}}$.

\begin{figure}
\includegraphics[angle=270,scale=0.5]{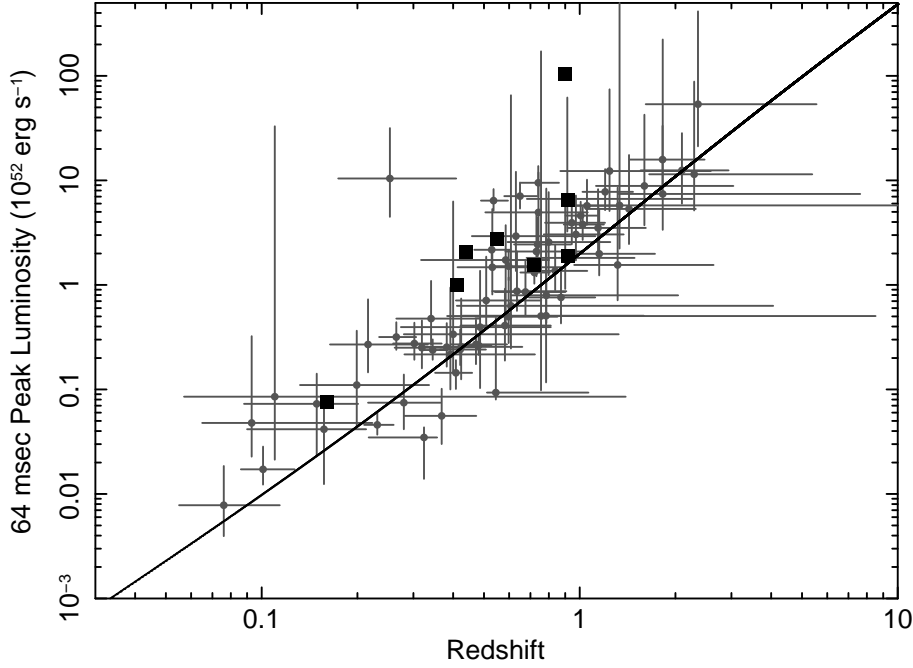}
\caption{The redshift distribution of SGRBs estimated by 
the $E_{p}$--luminosity correlation by \citep{tsutsui13}.
The solid squares are known redshift samples, and the solid
circles are ones of pseudo redshift. The solid line is the
flux limit of $4 \times 10^{-6}~{\rm erg~cm^{-2}s^{-1}}$
\label{fig1}}
\end{figure}

We succeeded in calculating all pseudo redshifts for 72 events.
In figure \ref{fig1}, we show the data distribution on the plane of 
redshift and 64ms peak luminosity. The filled squares and circles are 
known redshift samples with precious spectral parameters 
\citep[secure SGRBs by][]{tsutsui13} and pseudo redshift samples, 
respectively. The error of pseudo redshift is mainly caused by 
the statistical uncertainty of $E_{p}$, and the one of luminosity 
depends on the estimated redshift. The solid line is caused by 
the flux limit which must pass just close to the lowest and 
the highest data point because of a demand of our method to 
estimate the SGRB rate and luminosity function. If not, there is 
a possibility that the algorithm recognize unmeaning stronger 
luminosity evolution because of the lack of data around the flux limit. 
In this analyses, we set the flux limit of
$4 \times 10^{-6}~{\rm erg~cm^{-2}s^{-1}}$ to hold the number of 
data as many as possible.

\begin{figure}
\includegraphics[angle=270,scale=0.5]{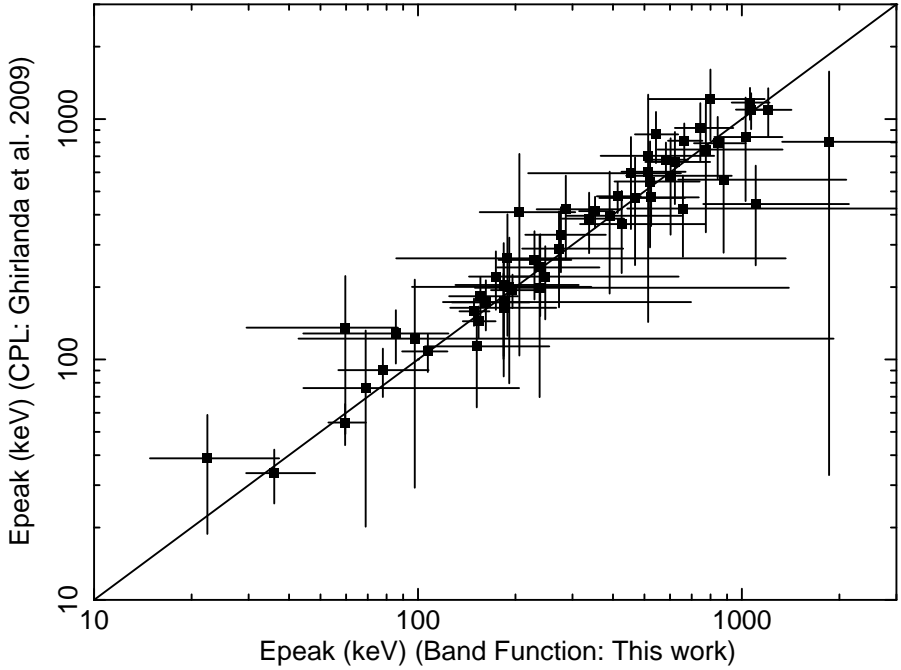}
\caption{The correlation between the $E_{p}$ values of this work 
(Band function) and the ones of \citet{ghirlanda09} (CPL function). 
The solid line is an equivalent line. Both results  strongly 
correlate with each other, but our results are slightly smaller than 
the one of \citet{ghirlanda09} for almost all samples. 
This trend comes from the different model function of spectral 
analyses as \citet{kaneko06} pointed out.
\label{fig2}}
\end{figure}

In figure~\ref{fig2}, we show the correlation between the 
$E_{p}$ value of this work (Band function) and the ones of 
\citet{ghirlanda09} (cutoff power-law function: CPL).
We confirmed both results strongly correlate with each other 
while our $E_{p}$ is slightly smaller than their results. 
This result is recognized as the difference of model function as 
\citet{kaneko06} mentioned. The cutoff power-law tends to
have larger $E_p$ than the Band function.

\begin{figure}
\includegraphics[angle=0,scale=1.4]{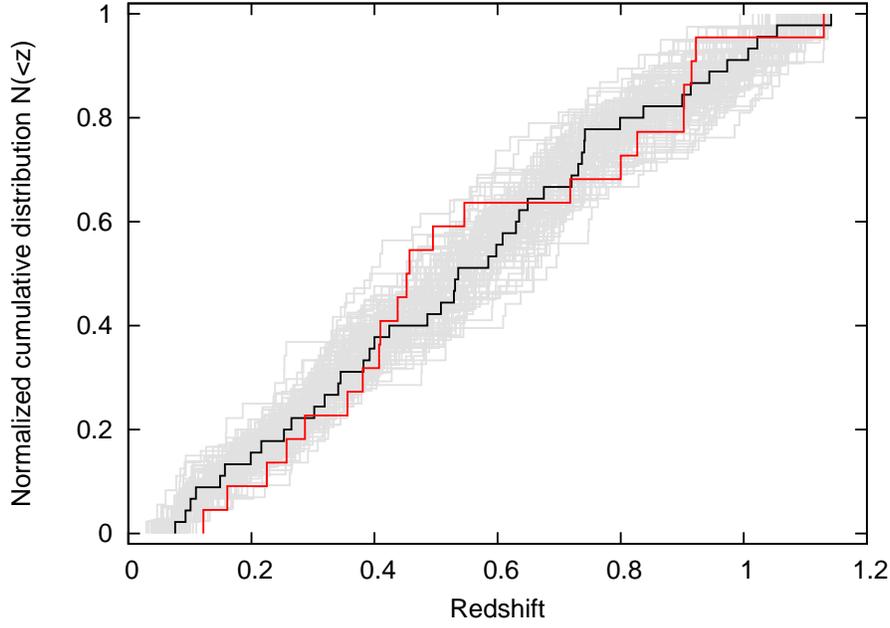}
\caption{The cumulative redshift distribution of SGRBs up to $z=1.14$. 
The black and the red solid lines are for 45 BATSE SGRBs in this paper
and 22 known redshift samples observed by HETE-2 and  {\it Swift}/BAT, 
respectively. The gray solid lines behind them show possible error 
regions estimated by 100 Monte Carlo simulations. We can see the good 
agreement of red, black and gray lines in the entire region.
Kolmogorv-Smirnov test between black and red lines shows that 
the probability that the two curves arise from different distribution
is 79.4~\%, and the error region shown in gray lines covers the 
red line. This strongly suggests that the $E_p$--$L_p$ correlation for 
SGRB \citep{tsutsui13} is a good distance indicator.
\label{fig3}}
\end{figure}

To confirm if our redshift determination is consistent with
one of the known redshift SGRBs, we compared the cumulative
redshift distributions of both samples. In figure~\ref{fig3},
we show the cumulative redshift distribution of 22 observed SGRBs
of $z \le 1.13$ (red) \citep{fong13a} and our 45 BATSE SGRBs 
brighter than the flux limit and with the pseudo redshift of 
$z \le 1.14$ (black) in figure~\ref{fig1}, respectively.
\footnote{In Table 3 of \cite{fong13a}, 37 SGRBs are listed.
However 11~SGRBs have either no firm redshift information,
for example, two redshift candidates or only upper/lower limits
of the redshift. Moreover we removed three possible host-less
SGRBs because their redshift is measured by the absorption lines
in the optical afterglow and they may be smaller than the real redshift.
We removed the most distant SGRB of $z=2.609$ to keep the shape of
cumulative distribution. Finally, we use only 22 SGRBs.}
The reason we set upper bound of the redshift comes from the small
number (only one) of the known redshift SGRB larger than $z=1.13$.

We performed the Kolmogorv-Smirnov test between the red and black 
lines in figure~\ref{fig3}, and it shows the probability of 
null-hypothesis is 79.4~\%. Moreover we estimate possible error
region of the cumulative distribution of 45 pseudo redshift samples.
As shown in figure~\ref{fig1}, the estimated redshifts have errors
mainly come from $E_p$ errors, so we performed 100 Monte Carlo 
simulations for each point and estimated their cumulative redshift 
distributions. The results are also shown as gray lines 
in figure~\ref{fig3}, and we can see the error region well contains 
the observed distribution (red line). Therefore we conclude
our estimated redshift distribution is the almost same distribution 
of observed one, and the $E_p$--$L_p$ correlation for SGRBs 
\citep{tsutsui13} is a good distance indicator. Hereafter we use 
53 SGRBs above the flux truncation of 
$4 \times 10^{-6}~{\rm erg~cm^{-2}s^{-1}}$ with maximum redshift 
$z=2.2$ to estimate the SGRB formation rate in the next section.

\subsection{Methodology}
In general, luminosity function can be written as
$\Psi(L,z) = \rho(z) \phi(L/g_k(z), \alpha_s)/g_{k}(z)$.
Here we named $\rho(z)$, $\phi(L/g_{k}(z), \alpha_s)$, and $g_{k}(z)$ 
are the SGRB formation rate, the local luminosity function, 
and the luminosity evolution, respectively. The parameter $\alpha_s$
means the shape of luminosity function, but we ignore this effect
because of the limited number of samples. The goal of this analysis
is to estimate the SGRB rate $\rho$ as a function of only $z$, 
and the local luminosity function $\phi(L/g_{k}(z))$ 
after removing the luminosity evolution effect.

The statistical problem to estimate the true SGRB formation rate 
and luminosity functions is how to deal the data set truncated 
by the flux limit. In many cases, assuming some parametric forms 
(model functions) for the luminosity function and redshift 
distribution, all parameters are simultaneously estimated to 
fit the data distribution of the flux limited samples.
However, if we use the model function far from the true distribution,
we may obtain unrealistic solutions for each parameters. 
Especially, we have little knowledge about the functional form 
of SGRB formation rate and it may be different from the general 
star formation rate. Therefore it is preferable to use 
a non-parametric method.

In this paper, we used a non-parametric method by 
\citet{lyndenbell1971, efron1992, petrosian1993, maloney1999}
developed to estimate the redshift distribution of distant Quasars. 
This method is also used in the long GRBs
\citep[e.g.][]{lloyd2001, yonetoku04, dainotti2008}.
The details of methodologies are found in these papers so that we
briefly summarized the thread of data analyses to estimate the 
luminosity function and the SGRB formation rate independently. 
In this work we follow the notations and terminologies by 
\citet{yonetoku04} to identify the best luminosity function 
distribution of $\Psi(L,z)$; see their section 4.

\subsection{SGRB Formation Rate}
First of all, we estimate the correlation between the redshift and 
the luminosity (luminosity evolution) with the assumption of 
the functional form of $g_{k}(z)=(1+z)^{k}$. Then we searched 
the appropriate $k$-value which gives the data distribution on
$(z, L/g_{k}(z))$ plane has no correlation between them.
Then we calculated the $\tau$-statistical value (similar to Kendall 
$\tau$ rank correlation coefficient) to measure the correlation 
degree for the flux truncated data. When the $\tau$ value is zero, 
it means that the combined luminosity $L/g_{k}(z)$ is independent
of the redshift $z$ (no luminosity evolution). We estimated 
$k = 3.3^{+1.7}_{-3.7}$ with 1~$\sigma$ uncertainty, so we can say
there is no obvious luminosity evolution ($g_{k}(z) \equiv 1$).

Next, we can separately calculate the local luminosity function for 
$L/g_{k}(z)$, i.e. $L$ for $g_{k}(z) = 1$, and the SGRB formation 
rate as a function of redshift with non-parametric method. 
We have already removed the effect of luminosity evolution, 
a unique formula for the luminosity function can be adopted 
for all redshift range. Then, we can easily estimate the number 
of events lower than the flux limit. In the same way, 
we can also estimate the SGRB formation rate. 

\begin{figure}
\includegraphics[angle=270,scale=0.5]{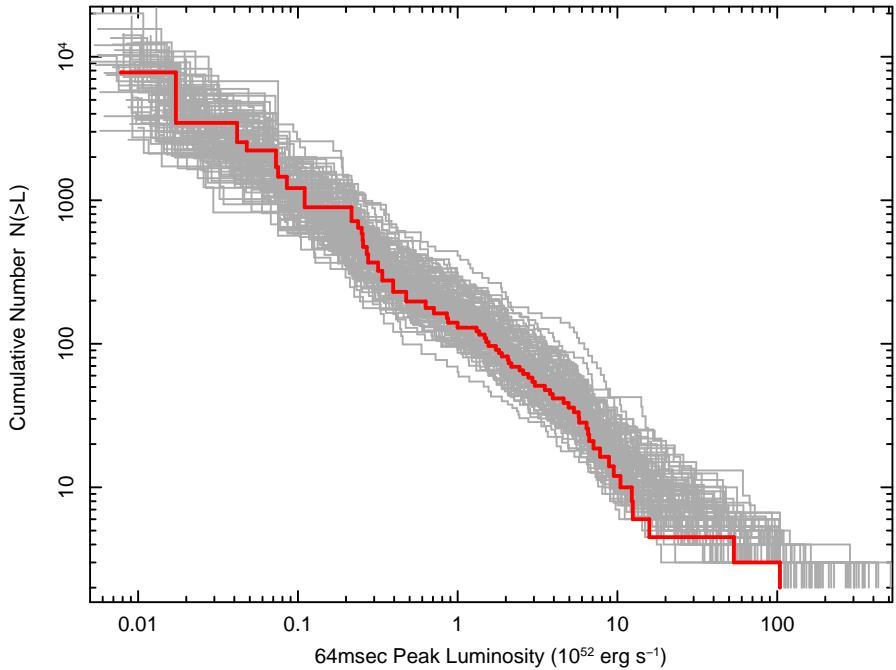}\\
\caption{The luminosity function of SGRBs estimated from data
distribution of figure~\ref{fig1}. The red solid line shows 
one of the best estimation, and 100 gray lines are possible 
error region estimated by the Monte Carlo simulations. 
 We can approximately describe it as a simple power-law function 
with the index of $-1$, and no obvious break has been found.
\label{fig4}}
\end{figure}

In figure~\ref{fig4}, we show the cumulative luminosity function 
of $L/g_{k}(z)$. The red line is the best estimate with the pseudo
redshift, and gray lines are results by 100 Monte Carlo simulations
as previously shown. For long GRBs, several authors reported 
the luminosity function can be described as broken power-law
\citep[e.g.][]{yonetoku04}. However, in this analysis for SGRBs, 
we can not find obvious break structure in figure~\ref{fig4}. 
We adopt a simple power-law function, and obtained the best fit 
index of $-0.84_{-0.09}^{+0.07}$ between the luminosity range of 
$10^{51}$ and $10^{53}~{\rm erg~s^{-1}}$. We can say the luminosity 
function is consistent with the pure unbroken power-law for 
$L>10^{50}~{\rm erg~s^{-1}}$.

\begin{figure}
\includegraphics[angle=270,scale=0.5]{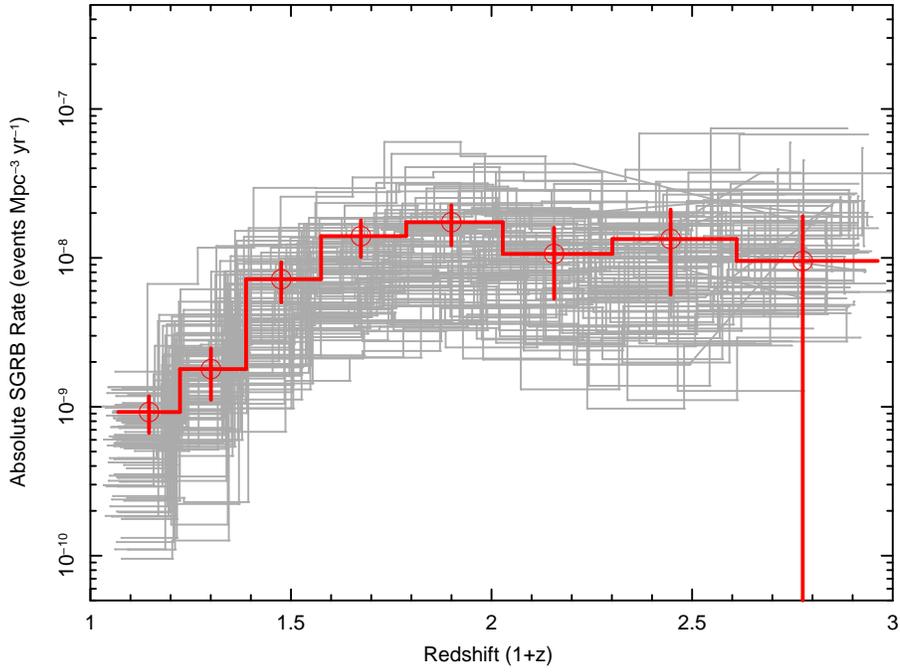}\\
\caption{Absolute formation rate of SGRBs estimated from data
distribution of figure~\ref{fig1}. Again the red line is the 
best estimation and 100 gray lines are ones from Monte Carlo
 simulations. The local event rate at $z=0$ is 
$\rho_{SGRB}(0)= 6.3^{+3.1}_{-3.9} \times 10^{-10}~
{\rm events~Mpc^{-3}yr^{-1}}$.
\label{fig5}}
\end{figure}

In figure~\ref{fig5}, we show the SGRB formation rate per comoving 
volume and the proper time as a function of $(1+z)$. Again, 
the red line is the best estimate with the pseudo redshift, 
and gray lines are results by 100 Monte Carlo simulations.
Here, we used the BATSE's effective observation period of 4.4~years 
as already explained in section~\ref{sec:data}. This SGRB rate is 
calculated for the events with the peak luminosity of 
$L > 10^{50}~{\rm erg~s^{-1}}$ in observer's frame. 
The functional form can be described as
\begin{eqnarray}\label{eq:sgrbrate}
\rho_{\rm SGRB}(z) \propto \left\{ 
\begin{array}{ll}
        (1+z)^{6.0 \pm 1.7}  & {\rm for}~(1+z) < 1.67,\\
        {\rm const.} & {\rm for}~(1+z) \ge 1.67,
\end{array}
\right.
\end{eqnarray}
in units of ${\rm events~Mpc^{-3}yr^{-1}}$. 
The local minimum event rate at $z=0$ is 
$\rho_{\rm SGRB}(0)=6.3_{-3.9}^{+3.1} \times 10^{-10}~
{\rm events~Mpc^{-3}yr^{-1}}$. 
Here, in this figure, we assume that the radiation of SGRB's prompt 
emission is isotropic, and we do not include any geometrical correction 
for the jet opening angle. In this analysis, we treated SGRB samples 
with the observed flux larger than 
$4 \times 10^{-6}~{\rm erg~cm^{-2}s^{-1}}$ and dimmer SGRBs are not 
included. Therefore the SGRB formation rate estimated here is 
regarded as the minimum value. 

Let us assume that progenitor of SGRBs is the merging 
neutron star-neutron star (NS-NS) binary here. 
\cite{Kalogera04a, Kalogera04b} obtained the probability function 
of the rate of the merging NS-NS binary taking into account 
the observed NS-NS binary, the beam factor of the pulsar, 
pulsar search time, the sensitivity and so on. 
They obtained a merging rate of 
$R_{\rm m}=10^{-7}-4\times 10^{-6}~{\rm events~y^{-1}Mpc^{-3}}$ 
with 99\% confidence level. 
\footnote{There are errors in \cite{Kalogera04a} so that the correct 
one is given in \cite{Kalogera04b}}. 
While \cite{Kim10}  analyzed the pulsar beaming effect with  newly found NS-NS binary to obtain the merger rate
of NS-NS binary as  $R_{\rm m}=9\times 10^{-7}{\rm events~y^{-1}Mpc^{-3}}$ which is within the
99\% confidence level of \citep{Kalogera04a, Kalogera04b}. For a review of various estimation of the merging rate , see  \cite{abadie10}.
From   $\rho_{\rm SGRB}(0)$ and $R_{\rm m}$, under the
hypothesis that every NS-NS merger produces a short GRB, we infer that any beamed emission must be confined
to a cone with opening angle greater
than $\theta_j^{\rm min}$  determined by
\begin{equation}
1 - \cos \theta_j^{\rm min} = \frac{\rho_{\rm SGRB(0)}}{R_{\rm m}}.
\end{equation}
Then we estimated $\theta_j^{\rm min}=0.6^\circ$--$7.8^\circ$.

\section{Discussion} \label{discussion}
\label{sec:discussion}

Long GRBs(LGRBs) are believed to be caused by relativistic jets since 
the break of the afterglow light curves are seen for many LGRBs. 
The typical example is GRB990510 which shows achromatic break of 
the afterglow light curve \citep{Harrison1999}. The physical reason 
for the achromatic break of the light curve comes from the jet dynamics 
after the relativistic $\Gamma$ factor of the jet becomes smaller 
than the inverse of the jet opening angle \citep{Rhoads1999}. 
If SGRBs are also caused by relativistic jets, we can expect a jet 
break similar to that of LGRBs. However, so far, only GRB~130603B 
has multi-wavelength data from radio to X-ray to confirm 
the achromatic jet break of the afterglow. \citet{fong13b} determined 
the jet opening angle of GRB~130603B as $4^\circ - 8^\circ$, 
where the ambiguity comes from the uncertainty in the kinetic energy 
of the jet and the ambient gas density. Very interestingly, 
the opening angle of GRB~130603B is compatible with the minimum jet 
opening angle derived in the previous section.

For GRB~051221, the steepening of the afterglow light curve was observed
in X-ray band at $t \sim 5$ days. If we identify this steepening as the
jet break, the jet opening angle is determined as $5.7^\circ -7.3^\circ$ 
\citep{soderberg06}. For GRB111020A, \citet{fong12} argued a significant
break at $t \sim 2$ days in X-ray band and obtained the jet opening
angle of $3^\circ - 8^\circ$. For GRB090426, \citet{nicuesa11} obtained 
the jet opening angle of $4.4^\circ$ from a break at $t = 0.4$ day. 
The other jet opening angle information from SGRBs is lower limits 
(see section 8.4 of recent review by \citet{berger13}) so that 
we use only these four estimations of the jet opening angles. 
A simple average of these four angles is $\sim 6^\circ$. 
Taking this value, then, the event rate of SGRB including the off-axis 
ones becomes $\rho_{SGRB, all}^{min}(0)= 1.15_{-0.71}^{+0.57}\times 10^{-7}~{\rm events~Mpc^{-3}yr^{-1}}$.

Now let us assume that the central engine of SGRB is a coalescence of 
a neutron star-neutron star (NS-NS) binary. Then, we can obtain 
an estimate of the event rate of gravitational-wave detection. 
The detectable range of KAGRA, adv-LIGO, adv-VIRGO and GEO network is 
$\rm \sim 200~Mpc$ so that the minimum gravitational-wave detection 
rate is obtained as $\rm 3.8_{-2.2}^{+1.8} ~events~y^{-1}$. This is an estimate 
independent of the one based on pulsar observations 
\citep{Kalogera04a, Kalogera04b}. On the other hand, if the central 
engine of SGRBs is a coalescence of a black hole-neutron star (BH-NS) 
binary with masses, say $M_{\rm BH}= 10 M_{\odot}$ and 
$M_{\rm NS}=1.4 M_{\odot}$, the detectable range will become 
$\sim 3.4$ times larger because it is proportional to 
$M_{\rm chirp}^{5/6}$, where the chirp mass $M_{\rm chirp}$ of 
a binary is defined by 
$M_{\rm chirp} = (M_1M_2)^{3/5}/(M_1+M_2)^{1/5}$ \citep{Seto2001} 
where $M_1$ and $M_2$ are the masses of each compact object. 
In this case, the detection rate will be $\rm 146_{-83}^{+71}~events~y^{-1}$ .

In the above estimation, we used the BATSE SGRBs with a flux larger 
than $4 \times 10^{-6}~{\rm erg~cm^{-2}s^{-1}}$ which is shown by 
the solid line in figure~\ref{fig1}.  As is seen in figure~\ref{fig1}, 
there are 17 events below the flux limit and the event with the lowest 
fluxes located a factor of $\sim 4$ below the solid line. 
If we lower the flux limit by a factor of 4, noting that the cumulative 
luminosity function is proportional to $L^{-1}$ as seen in 
figure~\ref{fig4}, we can expect that the event rate can be a factor 
$\sim 4$ larger. Then the gravitational-wave detection rate becomes 
$\rm 15.2_{-8.8}^{+7.2} (584_{-332}^{+284})~events~y^{-1}$  for NS-NS (NS-BH) binary, respectively.  
While \cite{abadie10}  suggested that the likely binary neutron-star detection rate for the Advanced LIGO-Virgo network will be 40 events per year, with a range between 0.4 and 400 per year in their review paper.

\citet{coward12} analyzed 14 {\it Swift} SGRBs taking into account the
ratio of the rate of BATSE to {\it Swift} SGRBs, the k-correction, the
maximum distance observed by a certain event, the jet opening angle and
the probability of the event being SGRB. They obtained the event rate of
$8_{-3}^{+5}\times 10^{-9}$--$1.1_{-0.47}^{+0.7}\times 10^{-6} ~{\rm
events~Mpc^{-3}yr^{-1}}$ for the case of isotropic emission and beamed 
emission, respectively. In our analysis the minimum event rate is 
 $\rho_{SGRB, all}^{min}(0)=1.15_{-0.71}^{+0.57}\times 10^{-7}~{\rm events~Mpc^{-3}yr^{-1}}$ after the
geometrical correction of beaming angle, while the realistic one is 
$4.60_{-2.84}^{+2.28}\times 10^{-7}~{\rm events~Mpc^{-3}yr^{-1}}$. 
Although both analysis used completely different methods, 
they are consistent with each other. For NS-NS binary, the detection
range of adv-LIGO will be $\rm \sim 100~Mpc$ in 2016-2017 
\citep{abadie10,aasi13} so that there is a good chance of the first 
gravitational wave detection around 2017.

\section*{Acknowledgments}

This work is supported in part by the Grant-in-Aid from the Ministry of 
Education, Culture, Sports, Science and Technology (MEXT) of Japan, 
No. 23540305, No. 24103006 (TN), No. 25103507, No. 25247038 (DY), 
No. 23740179, No. 24111710, and No. 24340048 (KT).

\clearpage

\end{document}